\documentclass[useAMS,usenatbib]{mnras}
\usepackage{natbib}
\usepackage{lscape}
\usepackage{graphicx}

\title[LMT/AzTEC observations of $\epsilon$ Eridani]{Deep LMT/AzTEC millimeter 
observations of $\epsilon$ Eridani and its surroundings}

\author[M. Chavez-Dagostino et al.]{
M. Chavez-Dagostino$^{1}$\thanks{E-mail: mchavez@inaoep.mx (MC)}
E. Bertone,$^{1}$
F. Cruz-Saenz de Miera,$^{1}$
J. P. Marshall,$^{2,3}$ \newauthor
G. W. Wilson,$^{4}$
D. S\'anchez-Arg\"uelles,$^{1}$
D. H. Hughes,$^{1}$
G. Kennedy,$^{5}$
O. Vega,$^{1}$ \newauthor
V. De la Luz,$^{6}$
W. R. F. Dent,$^{7}$
C. Eiroa,$^{8,9}$
A. G\'omez-Ruiz,$^{1}$ 
J. S. Greaves,$^{10}$ \newauthor
S. Lizano,$^{11}$
R. L\'opez-Valdivia,$^{1}$
E. Mamajek,$^{12}$ 
A. Monta\~na,$^{1}$
M. Olmedo,$^{1}$ \newauthor
I. Rodr\' iguez-Montoya,$^{1}$
F. P. Schloerb,$^{4}$ 
Min S. Yun,$^{4}$
J. A. Zavala,$^{1}$
M. Zeballos,$^{1}$
\\
$^{1}$Instituto Nacional de Astrof\'{i}sica Optica y 
Electr\'{o}nica Luis Enrique Erro \#1, CP 72840, Tonantzintla, Puebla, M\'exico\\
$^{2}$School of Physics, University of New South Wales, Sydney NSW 2052, Australia\\
$^{3}$Australian Centre for Astrobiology, University of New South Wales, Sydney, NSW 2052, Australia\\
$^{4}$Department of Astronomy, University of Massachusetts, Amherst, MA 01003, USA\\
$^{5}$Institute of Astronomy, University of Cambridge, Cambridge CB3 0HA, UK\\
$^{6}$SciESMEX, Instituto de Geof\'isica, Unidad Michoac\'an, Universidad Nacional 
Aut\'onoma de M\'exico, Antigua Carretera a P\'atzcuaro 8701, \\ Morelia, Michoac\'an, CP 58089, M\'exico\\
$^{7}$ALMA SCO, Alonso de C\'ordova 3107, Vitacura, Casilla 763 0355, Santiago, Chile\\
$^{8}$Departamento de F\'isica Te\'orica, C-XI, Facultad de Ciencias, Universidad Aut\'onoma de Madrid, 
   Canto Blanco 28049, Madrid, Spain\\
$^{9}$Astro-UAM, Unidad Asociada UAM - CSIC, Madrid, Spain\\
$^{10}$School of Physics and Astronomy, Cardiff University, CF24 3AA, UK\\
$^{11}$Instituto de Radio Astronom\'ia y Astrof\'isica, Universidad Nacional Aut\'onoma de M\'exico,  
Antigua Carretera a P\'atzcuaro 8701, \\ Morelia, Michoac\'an, CP 58089, M\'exico\\
$^{12}$University of Rochester, Department of Physics  and Astronomy, Rochester, NY, 14627-0171, USA 
}

\begin{document}

\date{Accepted . Received; in original form}

\pagerange{\pageref{firstpage}--\pageref{lastpage}} \pubyear{2016}

\maketitle

\label{firstpage}

\begin{abstract}
$\epsilon$ Eridani is a nearby, young Sun-like star that hosts a ring of cool debris analogous to the solar 
system's  Edgeworth-Kuiper belt. Early observations 
at (sub-)mm wavelengths gave tentative evidence of the presence of inhomogeneities in the ring, 
which have been ascribed to the effect of a putative low eccentricity planet, orbiting close to the 
ring. The existence of these structures have been recently challenged by high resolution interferometric 
millimeter observations. Here we present the deepest single-dish image of $\epsilon$ Eridani  at 
millimeter wavelengths, obtained with the Large Millimeter Telescope Alfonso Serrano (LMT). 
The main goal of these LMT observations is to confirm (or refute) the presence of 
non-axisymmetric structure in the disk. The dusty ring is detected for the first time 
along its full projected elliptical shape. The radial extent of the ring is not spatially 
resolved and shows no evidence, to 
within the uncertainties, of dust density enhancements. Additional features of the 1.1~mm map 
are: (i) the presence of significant flux in the gap between the ring and the star, 
probably providing the first exo-solar evidence of 
Poynting-Robertson drag, (ii) an unambiguous detection of emission at the stellar position
with a flux significantly above that expected from $\epsilon$ Eridani's photosphere, 
and (iii) the identification of numerous unresolved sources which could correspond to 
background dusty star-forming galaxies.
\end{abstract}


\section{Introduction}

The circumstellar debris disks detected around mature, main sequence stars are a visible 
remnant of planet formation processes \citep{Backman93}. Composed of icy and rocky 
bodies ranging from micron-sized grains to kilometre-sized planetesimals, the presence of a disk 
is typically revealed through the detection of excess emission from the star at mid- and 
far-infrared wavelengths \citep{Wyatt08,Matthews14}. 

Recent surveys by the 
{\it Herschel} Space Observatory  \citep{Pilbratt10} have identified cool disks, analogs to the 
Edgeworth-Kuiper belt of our solar system, in around 20~$\pm$~2\% of Sun-like stars \citep{Eiroa13}. 
However, their detection rate depended on both the temperature of the host star and on the 
observing strategy of the space craft. For instance, \cite{Thureau14} found
an incidence of 30\% in A-type stars. The combination of exoplanet and debris disk surveys has 
provided evidence that planets are more common around stars that also host a debris disk \citep{Bryden13}, 
and revealed tentative correlations linking the presence of dust and planets with the properties of 
the host star \citep{Wyatt12,Maldonado12,Marshall14a,Moro-Martin15}. The presence of a 
planet around a host star can be revealed through its dynamical interaction 
with the debris disk which creates non-axisymmetric structures (clumps, warps, cavities, etc.) 
in the disk. Such structures led to the discovery of a giant planetary companion in the prototypical 
debris disk host $\beta$ Pictoris \citep{Lagrange10}.
Thermal emission from the micron sized dust grains dominates the observed flux of the disk at far-infrared
wavelengths, and exhibits typical temperatures of 30 to 80~K \citep{Morales11} and radial 
size scales of 10s to 100s of astronomical units (AU) (\citealp{Pawellek14}, \citealp[see also][]{Marshall14a}). 
Tracing the largest millimeter sized and coolest grains in the disk, which do not drift as 
far from their parent planetesimal belt under the action of radiation forces as the smaller 
micron sized grains \citep{Krivov08}, is vital to accurately determine the 
location of the dust-producing belt of planetesimals 
around the star \citep{Krivov10}. Such measurements are only possible with (sub-)millimeter
continuum imaging observations \citep[e.g.][]{Williams06,Nilsson10,Panic13}. Resolving the radial extent 
of the disk is fundamental in the modeling process as it directly constrains the orbital 
radius of the dust responsible for the observed emission, weakening inherent degeneracies between grain 
size and radial distance in those models reliant solely on the disk thermal emission derived from 
analyses of the spectral energy distribution \citep[SED;][]{Augereau99,Lebreton12,Ertel14,Marshall14b}.

$\epsilon$ Eridani  ("Ran", HR1084, HD22049, HIP16537)  is a relatively young 
(age=0.8~Gyr, \citealp{DiFolco04,mamajek2008};  
1.4~Gyr, \citealp{Bonfanti15}), nearby ($d=3.22~$pc) Sun-like (spectral class K2V) star. Its 
age and distance place it as the closest isolated star of this kind where we can study the 
early stages in the evolution of a planetary system analogous to the solar system.  
The star is host to a bright, extended, almost face-on debris disk, which ranks amongst 
the finest examples of these objects 
so far discovered \citep{Greaves98,Holland98}. Recent models \citep{Backman09,Reidemeister11}
suggest that the disk is comprised of up to four distinct components: two warm inner belts, 
a cold outer belt and an extended halo of small grains. In these models the dust 
in the warm components actually originates in the the cold belt and is transported to the 
inner regions through the Poynting-Robertson drag and stellar winds. Radial velocity analyses 
suggest the existence of two giant exoplanets in addition to the warm inner debris 
disk \citep{Hatzes00,Moran04}. These planets are inferred to be within a few AU of 
the star, however their existence still remains contentious due to the high level of activity 
of $\epsilon$ Eridani, making interpretation of the spectroscopic measurements 
difficult \citep{Zechmeister13}.

The cold outer belt of $\epsilon$ Eridani has been extensively studied  from far-infrared (FIR) 
to (sub-)mm wavelengths from the ground \citep{Greaves98,Schutz04,Greaves05,Backman09,
Lestrade15,MacGregor15} and from space \citep{Gillett86,Backman09,Greaves14}. These observations showed 
that the debris disk has a ring-like morphology and provided the first estimates of 
the basic physical properties of the ring such as radial extent, width and inclination.  
Early sub-millimeter observations conducted with the  SCUBA camera on the 
James Clerk Maxwell Telescope \citep[JCMT,][]{Greaves98,Greaves05} also suggested that the ring 
has a clumpy structure that has been interpreted as evidence of dynamical interaction 
between an unseen planetary companion and the debris belt \citep{Quillen02}. 
Substructures in the ring were also identified in {\it Herschel}/PACS images at 160~$\mu$m by 
\cite{Greaves14} who, after considering limb brightening effects due to inclination, obtained a 10\% 
flux residual when comparing the flux in the southern portion of the ring to that of the north. Additionally, 
recent deep (rms $\sim$ 0.8~mJy/beam) observations made with MAMBO on IRAM at 1.2~mm  
\citep{Lestrade15} appear to corroborate the disk inhomogeneities found 
by SCUBA, though with one of the prominent clumps present in the SCUBA image being 
absent in the MAMBO map. The time span (18 years since the first sub-mm detection) 
between the different observational data sets and the large proper motion of  
$\epsilon$ Eridani, 1$\arcsec$/yr, have enabled studies of the outer belt's structure and dynamics, 
identifying which clumps in the belt could be associated with the disk and which might correspond to 
background sources, and to look for positional changes of the belt structures over time.  
The disk orbital motion has been estimated 
to be of the order of 1$^{\circ}$/yr \citep{Greaves05} or three times as large \citep{Poulton06}. 
There are, however, contrasting results. Observations at 1.2~mm with the bolometer array 
SIMBA on the SEST telescope (at a depth of 2.2~mJy/beam rms) did not confirm the presence 
of substructure in the ring \citep{Schutz04}, in agreement with the very recent
interferometric map at 1.3~mm collected with the Submillimeter Array (SMA) by 
\cite{MacGregor15}. They found that a smooth ring model could explain their 
patchy high resolution image. There are other examples in which the presence of dust density 
enhancements based on early observations at long wavelengths have been questioned by
more recent high resolution imagery. Noteworthy is the case of the very prototypical object 
Vega for which SCUBA observations in the sub-mm revealed bright blobs \citep{Holland98}, 
but whose detection was later disputed \citep[e.~g.][]{Hughes12}.

Motivated by the debated presence of structure along the ring around $\epsilon$ Eridani, 
its potential correlation with an inferred planet orbiting close to the inner edge of 
the ring, and the possibility of measuring the orbital motion of dust 
enhancements within the ring, we conducted deep continuum observations 
at 1.1~mm with the AzTEC instrument on the Large Millimeter Telescope Alfonso 
Serrano\footnote{www.lmtgtm.org} (LMT). 
In Section 2, we describe the LMT observations and reduction techniques. In Section 
3, the global observational morphology of the ring is presented. Section 4 is devoted to 
the detailed modeling of the ring structure. In Section 5, we briefly discuss the 
spectra of the components of the system. Section 6 provides additional comments 
on the flux detected at the stellar position, and in Section 7 we briefly comment 
on the background sources towards the $\epsilon$ Eridani's system. The concluding 
remarks are given in Section 8.

\section{Observations and data reduction}

Observations with the 1.1~mm continuum camera AzTEC 
\citep{Wilson08} were conducted in November and December 2014 as part of the 
Early Science Phase-3 of the LMT, while the 50-m diameter telescope was operating in its 32-m aperture 
configuration. The telescope is located on top 
of the extinct volcano Sierra Negra, in the state of Puebla, Mexico, 
at an altitude of 4600~m above sea level. 
A total of 18.5 hours on source were devoted to the target 
under weather conditions that ranged from excellent to good 
($\tau_{225GHz}$ = 0.03--0.11). The field was observed with the AzTEC small-map observing 
mode which covers an area of about 7.5~arcmin$^{2}$. The point spread 
function (PSF) of the instrument in this configuration has a FWHM beam size of 
8.5$\arcsec$, however, filtering in the reduction process results in an effective 
resolution of 10.9$\arcsec$. We made observations of the quasar 0339-017 roughly every 
hour which bracketed our observations of the $\epsilon$ Eridani field.  The measured 
quasar pointing offsets (typically $<$ 5$\arcsec$) were then interpolated in time 
to remove any offset and drift in pointing from the science observations.

The raw data were reduced using the standard AzTEC analysis pipeline and 
analysis approach \citep{Scott08,Wilson08} but without 
applying the final Wiener filter that many AzTEC observations 
use for the optimal detection of point sources.  
Instead, the final unfiltered image was smoothed with a Gaussian filter 
with FWHM=6.8$\arcsec$.  The data were flux calibrated 
based on observations of the proto-planetary 
nebula CRL618 and the noise in the final image is estimated from 
jackknifed time streams created as described in \cite{Scott08}. 

\begin{figure*}
\centering
\begin{minipage}{180mm}
\includegraphics[width=160mm]{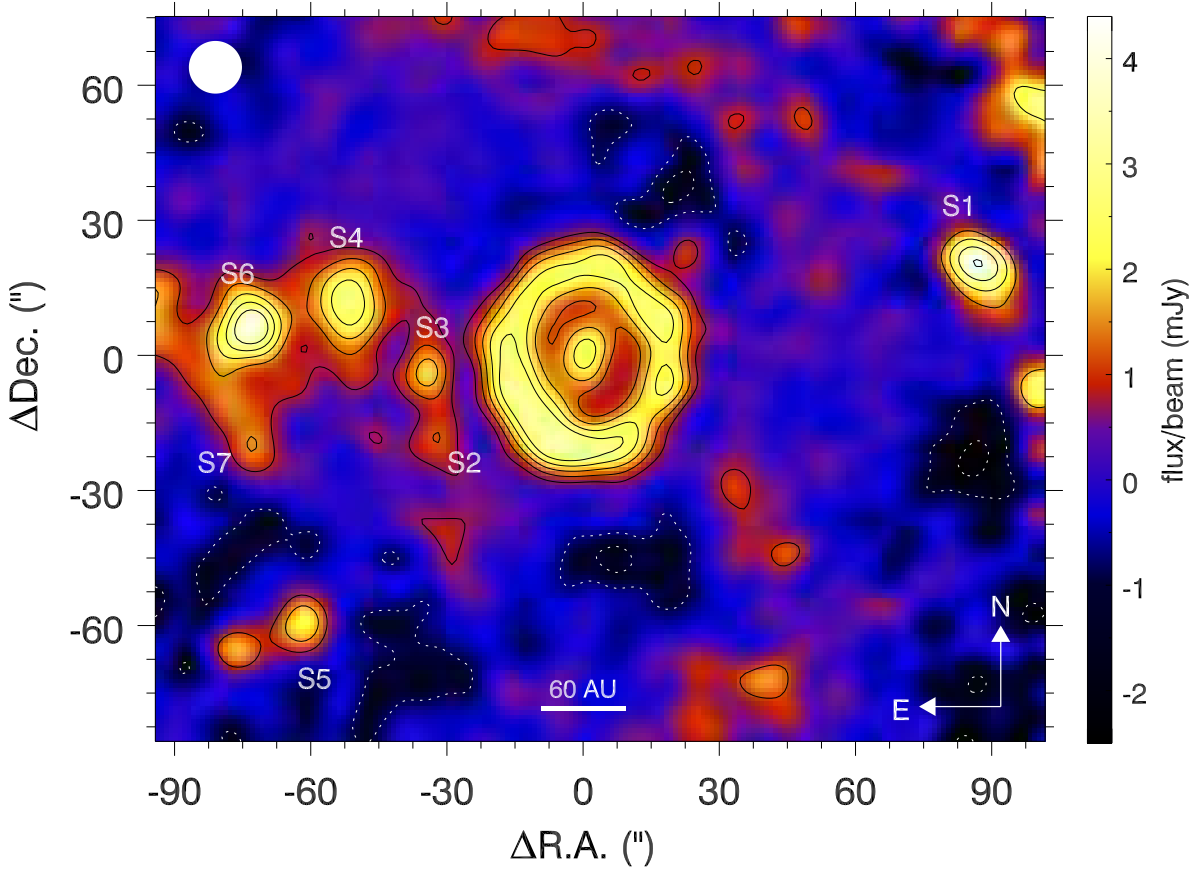}
\caption{1.1~mm LMT/AzTEC continuum map of $\epsilon$ Eridani. The outer ring is 
fully detected at a significance of $>$5.7$\sigma$. The central peak is 
detected at 7.5$\sigma$ and likely corresponds to the sum of three contributing 
agents: the stellar photosphere, the stellar upper atmosphere and an 
(or perhaps two) inner warm disk(s). As many as seven (S1-S7) background objects 
are detected in this map, of which, four have S/N$>$5. The source labeled 
S1 is the brightest with a 1.1~mm flux of 4.6$\pm$0.6~mJy. The smoothed beam size 
of 10.9$\arcsec$ FWHM is given by the white circle in the top-left. As a reference, 
we include the contour levels for S/N = -3.5, -2, 2, 3.5, 5.0, 6.5, 8.0.
\label{fig:aztec_map}}
\end{minipage}
\end{figure*}

Figure~\ref{fig:aztec_map} displays the final AzTEC 1.1~mm map which 
has an rms of 0.20~mJy/beam, about  ten times deeper than the SIMBA/SEST 
observations and four times deeper than those of MAMBO/IRAM.
The scale of the map is 1$\arcsec$/pixel which results in 93 arcsec$^{2}$/beam
for the smoothed beam size of 10.9$\arcsec$. This superb depth 
reveals three main interesting features in the map:  a well defined 
complete ring detected for the first time at millimeter wavelengths,  
a clearly detected central peak, and  numerous unresolved 
background sources, some of which were present in previous maps, in 
particular those from SCUBA/JCMT and {\it Herschel}/PACS and SPIRE instruments 
\citep{Poglitsch10,Griffin10}. Below, we describe in more detail 
these characteristics.

\section{Morphology of the dusty ring}

The ring is detected (at a significance ranging from 5.7 to 10.4$\sigma$) at all position 
angles and displays an almost perfect elliptical shape oriented along a north-south 
direction. Fitting an elliptical ring to the image results in  
a major axis of 20.0$\arcsec$ (or 64~AU) and the minor axis of 16.9$\arcsec$ (54~AU), 
which implies an inclination for a circular ring of about 32$^\circ$. This latter 
value is in agreement with the same 32$^\circ$ derived from {\it Herschel} data at 
160~$\mu$m \citep{Greaves14} and also, within their uncertainties, with the 
$\sim$17($\pm$14.2)$^\circ$ inclination estimated from the SMA map \citep{MacGregor15}.  The 
centroid of the ring matches the stellar position to within the expected pointing
uncertainties of the LMT, therefore we conclude 
that no offset is detected between the ring and $\epsilon$ Eridani.

The fact that we detect the full ring, clearly separated from a central peak, 
allows us to provide more observational constraints on the basic properties needed 
for the disk modeling, namely: the distance from the star (the ring radius $R$), the ring 
width (upper limit), and the dust properties obtained through the integrated flux density. 
In unresolved disks, whose dust characteristics can only be inferred through analysis of 
the SED, some of the above derived properties are, as mentioned before, degenerate. 

In addition to the ring radius $R$ and the inclination mentioned above, the ring appears unresolved 
in the radial direction (width $<$ 11$\arcsec$ or $\sim$35~AU) and has an integrated flux 
density, obtained through standard aperture photometry analysis, of 27.7$\pm$3.3~mJy, 
compatible with the values obtained at millimeter wavelengths with IRAM/MAMBO 
\citep{Lestrade15} and the SMA \citep{MacGregor15}. 

\begin{figure}
\includegraphics[width=84mm]{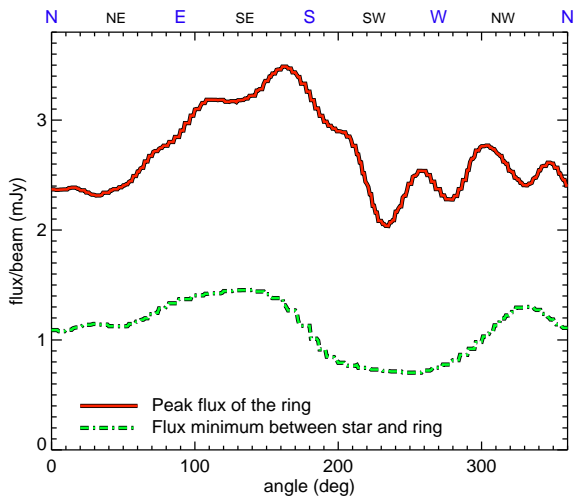}
\caption{Azimuthal flux distribution showing the variations of the flux maxima along the ring (continuous 
line). On average, the ring flux has a level of 2.4~mJy/beam, and the brightest arc of the ring is in 
the SE as can be seen in the map of Figure~\ref{fig:aztec_map}. The flux peak in this bright arc is 
$\sim$3.5mJy at about 160 degrees, measured counter-clockwise from north. The green dot-dashed 
line indicates the flux distribution in the gap which is also fully detected. The SE-NW ansae 
seen in Fig.\ref{fig:aztec_map} corresponds to the bumps in the gap distribution. The origin of 
this brightening is not known, but can plausibly be a reduction artifact.
\label{fig:az_peak}}
\end{figure}

\begin{figure}
\includegraphics[width=84mm]{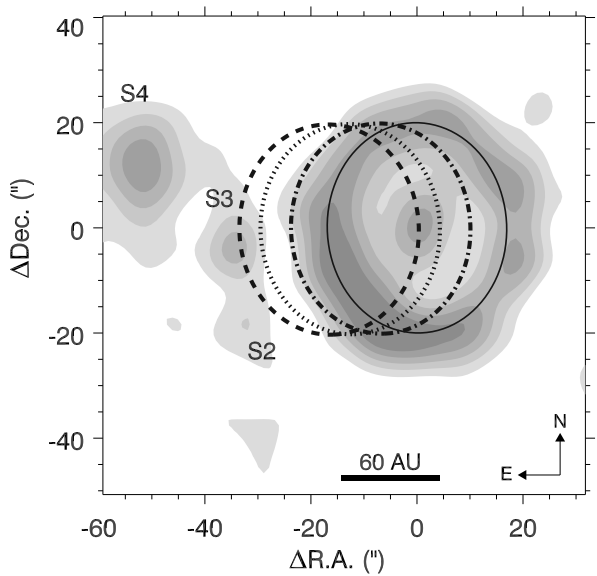}
\caption{Zoomed version of the 1.1~mm LMT/AzTEC continuum map of $\epsilon$ Eridani. 
The map is presented in gray scales that correspond to the contour 
levels of Figure~\ref{fig:aztec_map}. The ellipses denote the ring positions at
different epochs with respect to background point sources due to the proper motion of the 
system. The ellipses with thin solid, dot-dashed, dotted, and dashed lines show the location of the $\epsilon$ 
Eridani system at the epochs of the observations with AzTEC (2014.9), MAMBO (2007.9), and SCUBA (2002.0, 1997.9), 
respectively. Note that the source S3 almost coincides with the ring position at the 
first SCUBA observations of \citet{Greaves98}.}
\label{fig:aztec_rings}
\end{figure}

Morphologically, the ring in the LMT map shows a prominent 
extended brightening in the SE and three apparent clumps in the NW. This SE flux enhancement 
is evident from the azimuthal flux peak distribution depicted with the red line of 
Figure~\ref{fig:az_peak}. The maximum is at $\sim$160 degrees from the north counterclockwise 
reaching about 3.5~mJy/beam, 
which is $\sim$50\% above the lowest flux levels of the ring in the 0--40~deg and 
230--360~deg segments. This flux difference is certainly less notable than the factor 
of three reported in \citet[their Fig.~3]{Greaves05} and the factor of four in 
\cite{Lestrade15}. This SE bright arc 
is neither seen in the MAMBO map nor in SMA data. Conversely, the apparent clump in the 
SW in the SCUBA and MAMBO maps is not seen in the
LMT image.  While there are discrepancies, it is important to note that the published 
SCUBA data provided the most complete ring maps in the imaging data collected prior to these 
LMT observations,  and that the bright arc in the SE has been 
regarded as a real feature of the ring. Whilst a direct inspection of the positions of 
the flux maxima of this SE arc in the two SCUBA and the LMT images appear to indicate motion 
of substructures, the two sources east of the ring in the LMT map, in particular that 
labeled S3 in Figure~\ref{fig:aztec_map}, could conceivably explain the apparent flux enhancement 
in the SCUBA map of \cite{Greaves98} as its position agrees with that of the ring at the time 
when the SCUBA 
observations were carried out. Similarly, the suggested background sources as the origin of the 
flux brightening in the SW in the first SCUBA map could now well partially contribute to 
the LMT SE brightening. In Figure~\ref{fig:aztec_rings} we show the positions of the dust ring 
with respect to the stationary background sources at different epochs correspondent to previous 
(sub-)mm observations. 

To further verify the presence (or lack) of substructures along the ring, we modeled 
the $\epsilon$ Eridani LMT image using optically thin debris disk models as presented below.

\section{$\epsilon$ Eridani Modelling}

We modeled the emission from $\epsilon$ Eridani using a parameterized model of an
optically thin debris disk. This model has been described in detail in 
\cite{Wyatt99} and \cite{Kennedy12}. Basically,
a 3D distribution of dust surface area is generated, which can be viewed from any
direction to create synthetic images. The surface brightness of the images is calculated
by adding up the emission in individual cells along the line of sight, where each cell
emits like a blackbody at some temperature that is proportional to the distance to the
central star. For the resolved disk components we used a temperature law of $T_{\rm
res}=416/\sqrt{R}$~K, where $R$ is the disk radius in AU. Though we model the data at 
a single wavelength, this temperature was chosen to provide a reasonable extrapolation of the 
model to the photometric data at other wavelengths. For debris disks this model can
account for important effects such as brightening at disk ansae and brightness asymmetry
for non-axisymmetric disks. 

For a given model viewed at some orientation, a high resolution disk image is first
generated. The central stellar emission of 0.7~mJy (i.e. the expected photospheric flux,
see next section) is then added and the model is convolved with the LMT beam.
A $\chi^2$ goodness of fit metric is then computed within an
80$\times$80\arcsec~area shown in Figure~\ref{fig:models}, but only pixels (about 70\%) where the emission
from either the model or the image is significant are used \citep[for details see][]{Wyatt12}.
The variable background level means that simple least squares minimization does not necessarily
yield satisfactory results, so in most cases some by-hand intervention was needed to obtain a
smooth and continuous background. That is, even though models with $\chi^2$ values lower than those
presented are possible, they remove what could be astrophysical background behind
the disk, in particular at the SE side, and produce negative residuals in the NW.
While the $\chi^2$ is always several times higher than the number of degrees
of freedom (because our model of the sky near $\epsilon$ Eri is incomplete), visually better fitting
models do have lower $\chi^2$ values. As is commonly the case with models of low spatial-resolution data,
we do not explore all possible parameter space so do not claim that our models are unique, but that
they are reasonable interpretations of the LMT data.

The outer ring at $\sim$70 AU is clear in the LMT image, so the main goals of the modeling were
i) to determine whether the ring width was well constrained, ii) determine whether
additional emission above that expected from the star is present interior to the outer
ring, and iii) look for any evidence of azimuthal structure.

Additional issues for this modeling were the overall flux calibration and any flux (DC) offset,
and nearby background sources. 
We allowed for a small DC
offset in modeling the images, though it did not influence the results. Near the disk
three background sources were added, one corresponding to the blob in the NW, and two to the 
E and SE sources labelled S3 and S2 in Figure~\ref{fig:aztec_map}, respectively. As we note
below, there is the possibility of additional emission that exists behind the SE portion 
of the disk, which could plausibly be an extension of the diffuse emission to the 
E of the system.

\subsection{Ring width}
To address point  i) above, we constructed models with a range of ring widths with 
constant surface density centered at a distance of about 70 AU from the star. These models 
consider an inclination of 30$^\circ$ which, unlike previous observational results at 
millimeter wavelengths \citep[e.g.][]{MacGregor15}, is strongly constrained. The position
angle is $7^\circ$ from north to west (i.e. slightly west of north), so the geometry 
is consistent with previous results \citep{Greaves14,MacGregor15}. Models with 
arbitrarily narrow ring widths reproduced the data reasonably well, with $\chi^2_{\rm red}$ values of around 4.7 (where $\chi^2_{\rm red}$ is the reduced 
$\chi^2$), and these models also have a good 
fit to the dust emission interior to the main ring. We also constructed 
models with wide rings, finding that a width of up to 30~AU was acceptable, but 
for larger sizes the ring surface brightness becomes too low, i.e. for a 30 AU width 
$\chi^2_{\rm red}=5$. 
Some example models are shown in Figure~\ref{fig:models}. The lower two rows
of panels show the results for narrow (10~AU) and wide (30~AU) outer belts, both of which
produce reasonable results. Thus, we conclude that the LMT image does not place a lower
limit on the ring width and that it could be as wide as 30~AU, conclusions that compare
well with those of \cite{Lestrade15} and \cite{MacGregor15}.

\subsection{Non-stellar emission interior to the ring}
To address the second issue, we constructed models with an additional interior component. 
The inclusion of this extra component is motivated by two facts: 1) we expect a 
contribution of an excess above the photosphere from the star originating from the warm 
belt(s) and 2) the significant flux in the gap between the star and the ring we found 
in the LMT map (green dot-dashed line in Figure~\ref{fig:az_peak}).

\begin{figure*}
\begin{center}
\begin{tabular}{c}
\includegraphics[scale=0.4]{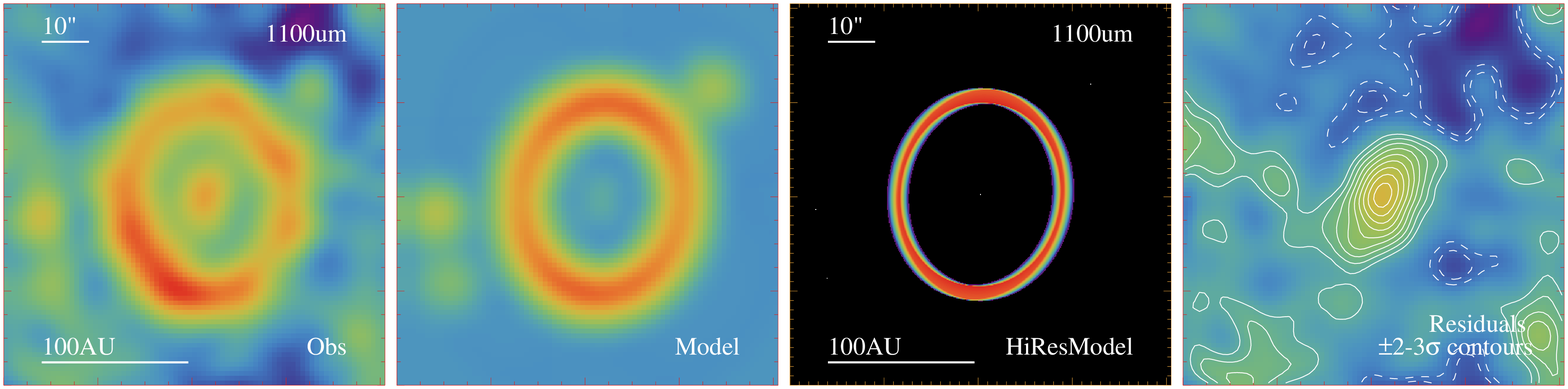} \\
\includegraphics[scale=0.4]{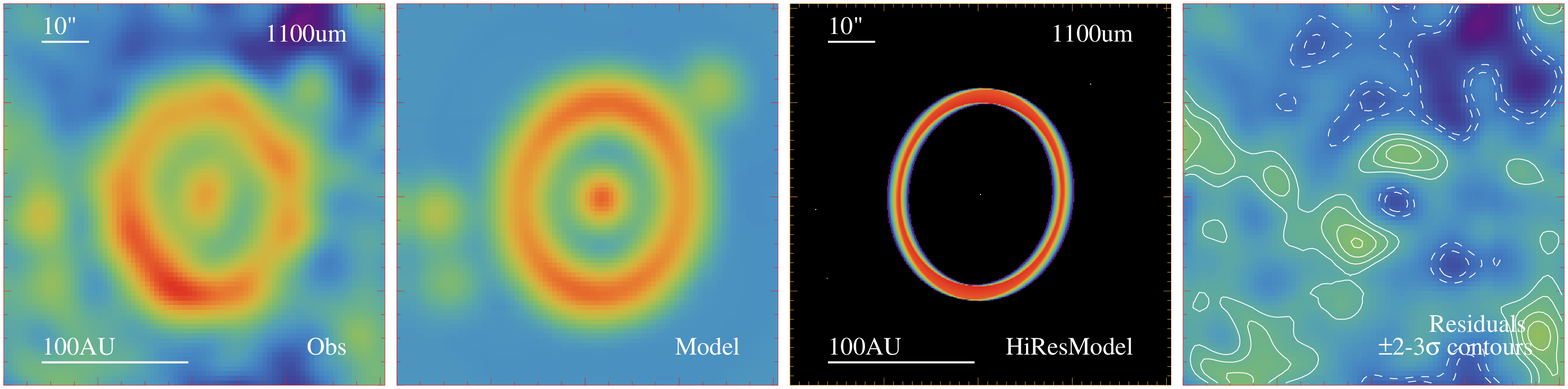} \\
\includegraphics[scale=0.4]{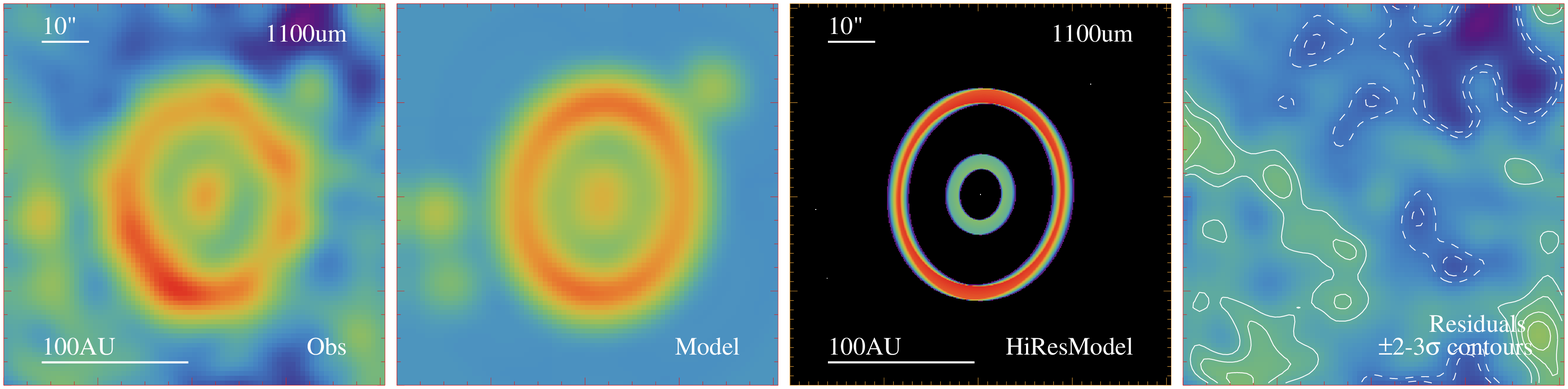} \\
\includegraphics[scale=0.4]{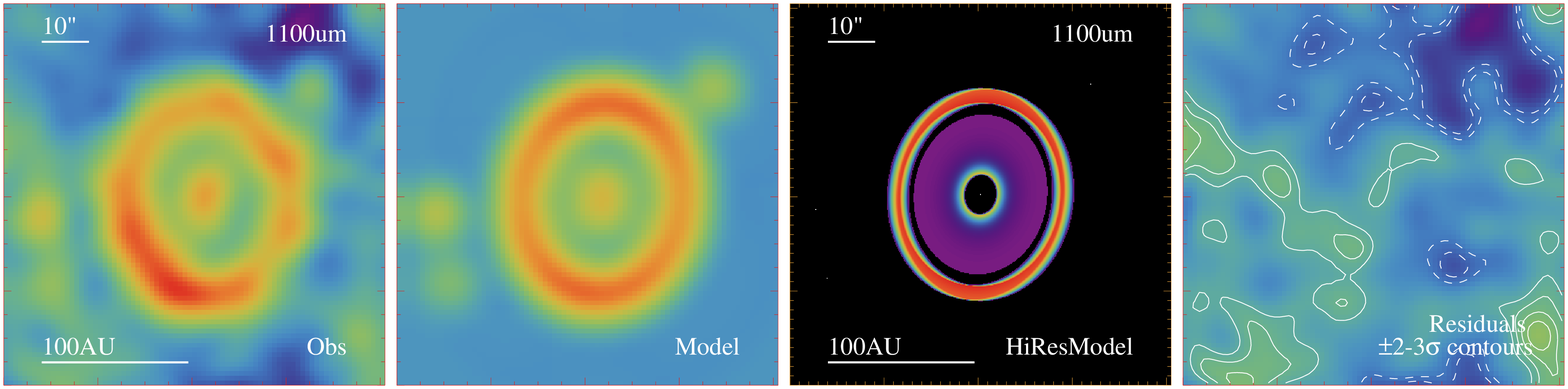} \\
\includegraphics[scale=0.4]{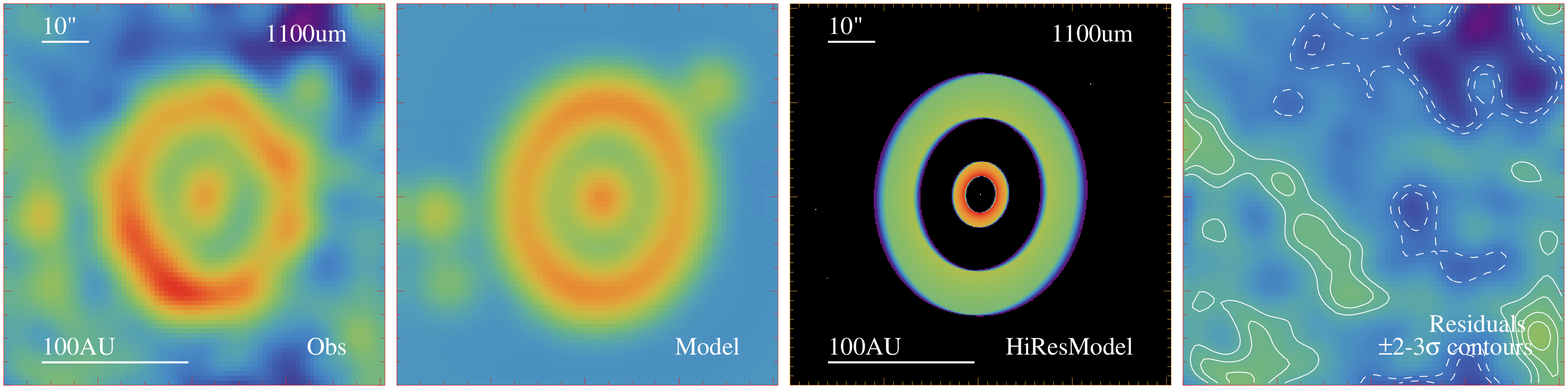} \\
\end{tabular}
\caption{Models of the LMT 1.1~mm image of $\epsilon$ Eridani. Each row is a different model, with the panels showing,
from left to right, the data, the model at the resolution of the 32-m LMT, the high resolution model, and the residuals. 
The residuals include contours in units of S/N from $\pm$2$\sigma$. The panels of the first and 
second rows show an outer ring centered at 68~AU and 10~AU wide, with the stellar photosphere 
(0.7~mJy) as the only contributor in the inner region (first row), and an artificial point source of 
2.7~mJy added at the stellar position (second row). The third row considers an outer ring as in the 
above panels and includes an inner ring 10~AU wide at 23~AU. In the fourth row the inner component 
is wider, extending from 14 to 63~AU, but follows a surface density power law with index=-3.5 in order 
to keep it concentrated. In the bottom row panels we considered a wider (30~AU) outer ring centered at 
69~AU, with a narrow (10~AU) inner component at 18~AU.
\label{fig:models}}
\end{center}
\end{figure*}

\begin{figure}
\includegraphics[width=84mm]{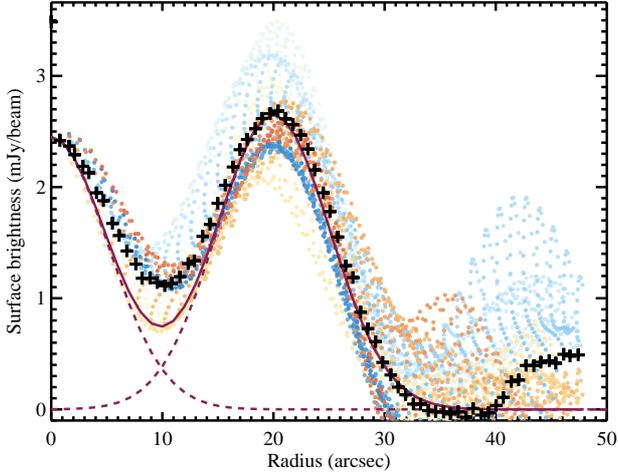}
\caption{Radial flux distribution of the $\epsilon$ Eridani system. The dotted lines are 
the Gaussian profiles fitted to the central peak and the ring. They both match the PSF of 11$\arcsec$, hence not 
resolved. The continuous line is the sum of these profiles. Colored dots represent the full set of flux 
points in each azimuthal direction, and the average observed fluxes are depicted with the "+" symbols. Note 
that significant diffuse flux is present in the gap at 10$\arcsec$ radius.
\label{fig:rprof}}
\end{figure}

The width and brightness of this component was varied to ascertain whether its existence was 
required by the data, and if so, the extent of this component. The above models that include only 
the outer belt leave significant residuals in the interior regions ($\chi^2_{\rm red}=8.2$), 
so we conclude that mm-wave emission exists in excess of the stellar photospheric emission
interior to the outer belt (e.g. top row of panels in Figure~\ref{fig:models}). A 2.7~mJy 
(i.e. 2~mJy above the photosphere) point source at the stellar position produces marginally 
satisfactory results ($\chi^2_{\rm red}=5.3$), but emission remains between the star
and the outer belt (second row of panels). The same conclusion can be reached by considering the 
radial flux 
distribution shown in Figure~\ref{fig:rprof}, where the outer belt
appears unresolved but the interior flux cannot be explained by simply adding a point
source at the star position. This flux is somewhat overestimated, as can be seen by
the negative residuals at the star position. We therefore favour models with extended
emission between the star and the outer belt. Models with interior emission included can 
reproduce the data fairly easily; both continuous emission from a flat surface density disk 
extending from 14 to 63~AU, and a narrow belt at 30~AU produced satisfactory results
($\chi^2_{\rm red} \approx 4.7$
(see the third and fourth rows in Figure~\ref{fig:models}). 
Thus, while we are confident that the interior emission exists, we cannot constrain from where 
this emission originates. In most models a small amount of emission was required at the stellar
position, but the level is degenerate with the structure of the emission between the star
and the outer belt.

\subsection{Azimuthal structure}
In all cases where a reasonable fit to the data was obtained there was no evidence for 
significant non-axisymmetric emission in the residuals. In most cases
emission remained in the SE part of the ring, but, given the extended structure seen to
the E of the system, attributing this residual emission to the $\epsilon$ Eridani system
is not well justified.

\begin{figure}
\includegraphics[width=84mm]{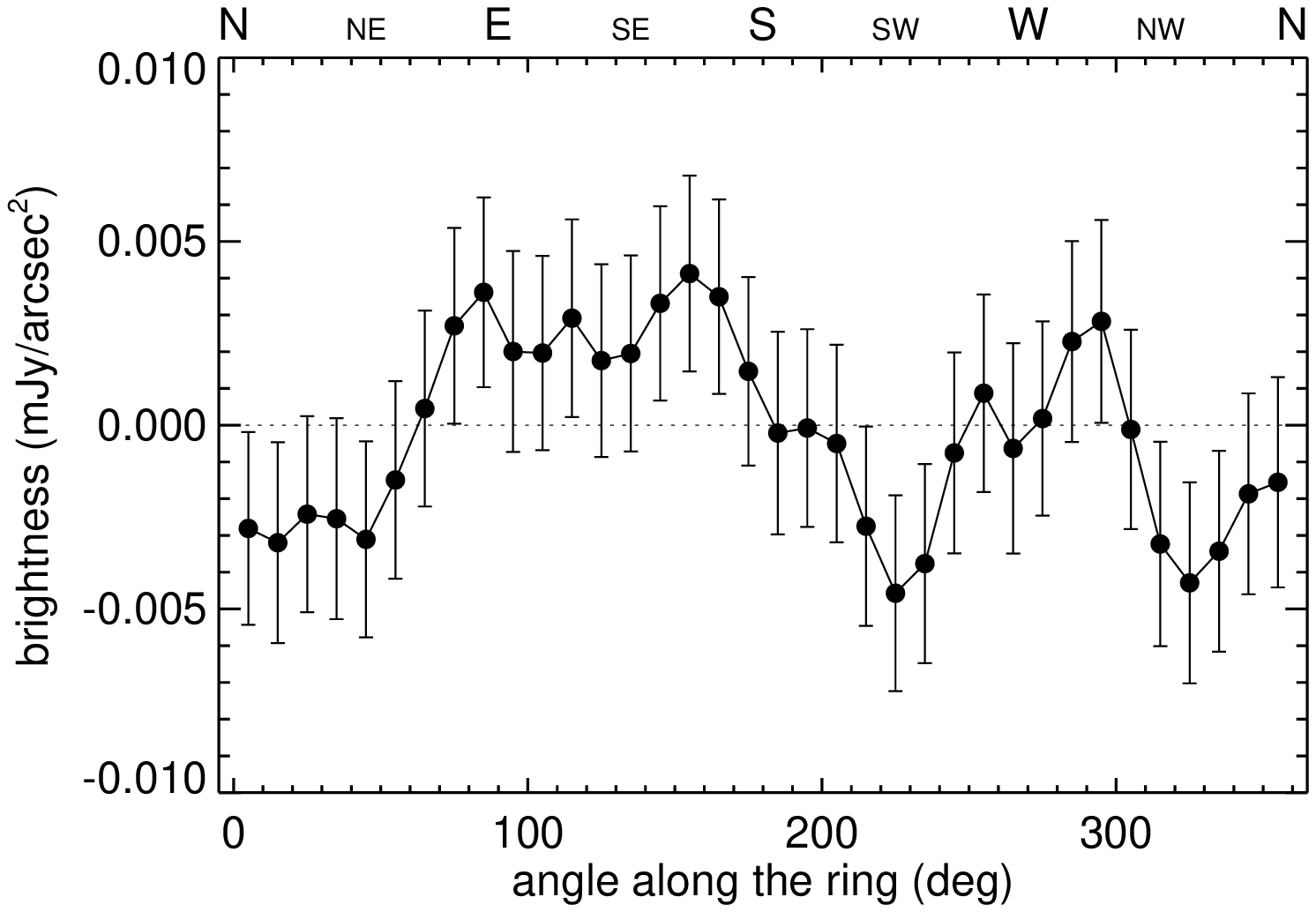}
\caption{Flux residual obtained by subtracting a well fitting model flux to that of the observed map. The residual 
indicates that, to within the uncertainties, the ring has a smooth morphology. A reduced $\chi^2$ of the data 
points results in 0.92, supporting lack of substructures.
\label{fig:az_res}}
\end{figure}

Complementarily, we carried out a similar analysis as that of 
\cite{MacGregor15}, namely, we calculated the residual emission by subtracting 
a best fit model of the ring to the integral of the observational flux calculated in 
sectors with a central angle of 10$^\circ$ extending from 10 to 28$\arcsec$ from 
the center. Note that the outer radius of the annulus is slightly smaller than in 
\cite{MacGregor15} because we wanted to avoid contamination from the faint NW 
source close to the ring. The azimuthal distribution of the residual flux is displayed 
in Figure~\ref{fig:az_res}. Note that the only potential feature detected in the NE quadrant of the 
SMA map is not present in our map. To within the uncertainties, the distribution shown in this 
figure indicates that the ring has a smooth structure, with perhaps the presence of 
two regions of low flux at azimuthal angles of $\sim$220 and $\sim$320 degrees. 

The various components in the best fit models have a 1.1~mm flux of $\sim$25~mJy for the
outer ring, in good agreement with the photometric result shown above,  and a total of 
5.5~mJy for the interior regions. A flux of 0.7~mJy is attributed to the stellar photosphere. 
For the model where the interior disk component is a narrow ring the flux is 3.4~mJy, and 
when it is extended the flux is 4.5~mJy. 

\section{The spectrum of the  $\epsilon$ Eridani system}

\begin{figure}
\includegraphics[width=84mm]{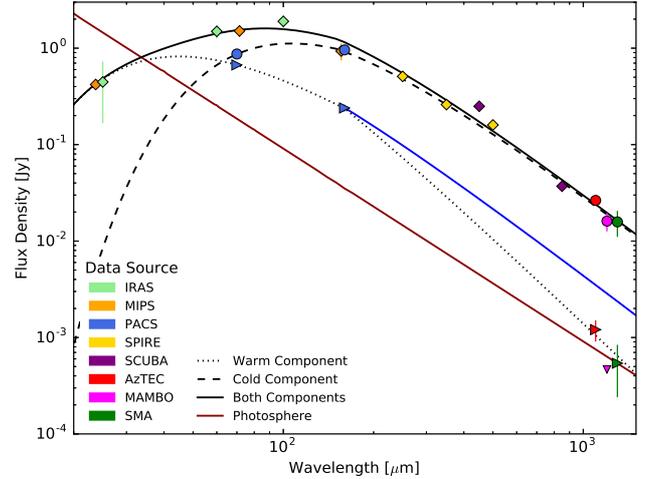}
\caption{Spectrum of $\epsilon$ Eridani.  Different lines styles represent 
the different components of the system: the photosphere (solid red), the outer belt (dashed), 
the inner component (dotted), and the summed contributions (thick solid). The instrumental origin 
of the ancillary data is indicated in the figure inset and the symbols stand for data that
contain fluxes for the inner component (triangles), the outer belt (circles), and
those of the whole system (diamonds). These data have been partially compiled by 
\citet{Greaves14} from \citet{Backman09} and \citet{Greaves05}. The outer belt spectrum has 
been constructed with the parameters $T$=48~K, $\lambda_{0}$=150~$\mu$m, and $\beta=0.4$, 
as in \citet{Greaves14}. 
The inner component, instead, represent the best fit of the four available points, two from {\it Herschel} 
PACS at 70 and 160~$\mu$m, the SMA data point at 1.3~mm, and the LMT/AzTEC flux at 1.1~mm. The best
fitting parameters assumming $\lambda_{0}$=150~$\mu$m are $T$=113~K and $\beta=1.0$. The solid blue 
line shows the long wavelength extension of the gray body emission for the inner component considering
an emissivity index of $\beta=0.4$. The two data points at long 
wavelengths (and the upper limit of MAMBO at 1.2~mm) provide meaningful constraints 
on the dust properties of the warm belt.
\label{fig:sed}}
\end{figure}

The infrared to millimeter spectrum of the $\epsilon$ Eridani system is depicted in Figure~\ref{fig:sed}. 
Three separate components are shown in this figure. The solid brown line 
represents the photosphere that was generated from an interpolation within the library of 
synthetic spectra of \cite{Castelli03}. These synthetic fluxes are calculated  up to 160~$\mu$m, 
so an extrapolation of 
the Rayleigh-Jeans tail was necessary to account for the flux at longer wavelengths. We 
have assumed the stellar parameters derived by 
\cite{Paletou15}: ($T_{\rm eff}$/$\log{g}$/[Fe/H])=(5034/4.51/+0.16). 
The dashed curve indicates the location of the modified blackbody curve for a temperature of
48~K, $\lambda_{0}=$150~$\mu$m, and an emissivity index $\beta=$0.4 \citep{Greaves14}. 
Ancillary data (see labels in the figure) are also plotted for illustrative purposes and 
no formal fit of the points was attempted. The data, in spite of some dispersion, 
are well represented by the above modified blackbody parameters. The dispersion can be partially 
attributable to the differing components that are included in the integrated flux densities 
calculated from distinct data sets. The dotted curve displays the best fit provided by the 
{\it Herschel}, SMA and LMT data for the inner component. The latter data point (1.3~mJy) 
includes only the unresolved central emission after subtracting the photospheric 
(and chromospheric, see next section) contribution considering $\lambda_{0}=$150~$\mu$m. 
The resulting best fit parameters of the warm component are $T=113$~K and $\beta=1.0$. 
The blue line shows the extension of the modified blackbody for the inner component using 
the parameters of \cite{Greaves14}. The solid thick black curve corresponds to the summed 
contributions of the photosphere, the inner flux peak, and outer belt.

With the available data, dust masses ($M_{\rm dust}$) can be calculated for the different 
components. As in \cite{Greaves98}, we consider two cases for the absorption coefficient; 
$k_{850\mu m}=$ 1.7 and 0.4 cm$^{2}$gr$^{-1}$ which at 1.1mm correspond to 
$k_{1.1mm}=(850/1100)^\beta \times k_{850\mu m}$cm$^{2}$g$^{-1}$. 
At a distance of 3.22~pc, $T=$48~K, and $\beta=0.4$ ($k_{1.1mm}=$ 1.53 and 0.36 cm$^{2}$g$^{-1}$), 
$M_{\rm dust}$ for the cold outer belt is 0.0035 and  0.015 M$_{\oplus}$, which are 
slightly smaller, but compatible with those reported by \cite{Greaves98}. 

For the inner component we conducted two calculations: a) we first 
considered the best fit parameters obtained above for the uresolved 
warm component and the transformed absorption coefficients $k_{1.1mm}=$ 1.31 and 
0.31 cm$^{2}$g$^{-1}$ for $\beta=1.0$. The resulting $M_{\rm dust}$ are 7.6$\times 10^{-5}$ 
and 0.0003~M$_{\oplus}$ for the two coefficient values, respectively; b) we assumed an inner component 
that includes both the unresolved component and a narrow warm disk with a total
flux of 3.4~mJy, in agreement with an emissivity index of $\beta=$0.4. In this case 
the dust mass is $M_{\rm dust}$=0.00017 and 0.0007 M$_{\oplus}$, for the high and low 
values of $k_{1.1mm}$, respectively. According to these results there is a similar amount
of warm dust (~0.0001~M$_{\oplus}$) very close to the star and in between the central 
unresolved emission and the external cold belt which could be, as mentioned before, 
evidence of material being tranported to the inner regions from the outer relatively 
massive debris ring.

\section{Additional comments on the central peak}

The unresolved emission from the central peak is detected at 7.5$\sigma$. The integrated flux in 
the center is 2.3$\pm$0.3~mJy and arises from several potential contributors, namely, the stellar 
atmosphere (photosphere and chromosphere) and the Rayleigh-Jeans tail of the warm dust 
component(s) identified at 70 and 160~$\mu$m 
with {\it Herschel}/PACS \citep{Greaves14}. As mentioned above, the expected photospheric 
contribution to the total flux at 1.1~mm is 0.7~mJy, implying a potential warm dust contribution 
of $\sim$1.6~mJy. The blue line in Figure~\ref{fig:sed} predicts an excess of $\sim$3.5~mJy at 
1.1~mm, which is nearly 
twice as much as that measured in our LMT map. In fact, the addition of the two extra data
points provided by the LMT and the SMA results in a much steeper $\beta$ index and 
suggests that $\lambda_{0}$ should remain close to the {\it Herschel}/PACS band at 160~$\mu$m. 
In addition, it has been found that active 
stars present a non-monotonic temperature variation towards the upper atmospheric layers, in a 
similar way the Sun displays a temperature 
increase towards low optical depths after reaching a minimum temperature in the atmosphere at 
about $\log{\tau}=-4$. This temperature gradient reversal is commonly evidenced by 
an excess in the ultraviolet regime but also as an atmospheric (sub-)mm-wave excess. In a 
recent study, 
\cite{Liseau15} found that $\alpha$ Cen B (K1 V) displays an excess at bands 9, 7 and 3 
(440~$\mu$m, 870~$\mu$m and 3~mm, respectively) of the Atacama Large Millimeter/submillimeter 
Array (ALMA).
Considering that $\alpha$ Cen B  and $\epsilon$ 
Eridani have similar spectral types and assuming that both stars have a similar degree of 
activity, as suggested by their far and mid-ultraviolet (1200--2500~\AA) continuum flux surplus, 
it is not unreasonable to expect a similar excess in $\epsilon$ Eridani.  The comparison of
observed and predicted fluxes of $\alpha$ Cen B indicates that in ALMA bands 7 and 3 the 
observed stellar fluxes are, respectively, 40 and 220\% higher than predicted for the 
photosphere. Based on observations conducted with the Australia Telescope Compact Array 
(ATCA) at 7~mm, \cite{MacGregor15} reported a flux excess of about a factor of three, which 
agrees with the trend of an increasing excess at longer wavelengths. These authors argue that no 
plausible inner disk scenario is able to explain the observed excess and ascribe it to a thermal 
origin in the upper atmosphere of $\epsilon$ Eridani. An interpolation of the bracketing 
ALMA points to the 1.1~mm band results in an excess of $\sim$50\%, which 
would imply a total stellar flux of $\sim$1~mJy for $\epsilon$ Eridani. This atmospheric 
value actually leaves 1.3~mJy at 1.1~mm originating from the unresolved warm  belt(s) 
and, therefore, its contribution would be about half of the measured central emission. 
Complementing the LMT and SMA data with measurements of the central source at other 
(sub-)mm bands will be very valuable to better assess the appropriate contributions of 
the sources of millimeter emission. These contributions will allow us to better constrain 
the dust properties of the warm belt and to understand the outer atmospheric structure of 
$\epsilon$ Eridani.

\section{Background sources}

As mentioned before, another important feature of our map is that, thanks to the depth and enhanced 
resolution achieved, we found numerous sources whose nature has been suggested, but never been 
previously investigated. In addition to the cold ring and the central peak, seven point sources 
were detected with a significance $\geq$3.5$\sigma$ and fluxes that range from 1.2 to 4.6~mJy 
(see Table~\ref{table:bkg_fluxes}).  These sources do not have near- or mid-IR
counterparts on either 2MASS or {\it WISE} images, but some are also present in previous {\it Spitzer} 
\citep[MIPS-70~$\mu$m;]{Rieke04}, {\it Herschel} (PACS and SPIRE at 70, 160, 250, 350 and 500~$\mu$m), 
and, particularly, in SCUBA and SCUBA-2 
images at 450 and 850~$\mu$m. For several of the sources (S2, S3, and S7) our map provides the 
first detection. It has been suggested by \cite{Greaves98,Greaves05} that these sources are probably 
members of the population of distant and heavily obscured sub-millimeter galaxies 
\citep[SMG;][]{smail1997,Hughes98}. Considering the area of the map, we expect $\sim$10 SMGs  
to be present in our map for a detection threshold of $S_{1.1mm}$ = 0.7~mJy (\citealp{Scott12}, 
\citealp*{Shimizu12}), which is consistent with the number of point sources 
detected in the AzTEC map.  It also agrees with the number counts derived in the very recent study
of \cite{fujimoto16} scaled to the 1.1~mm wavelength for a spectral index of 3. 
The analysis of these background sources 
is beyond the scope of this paper. We, nevertheless, would like to remark that photometric and 
molecular line studies are required to reveal the (distant or nearby) nature of the sources 
surrounding the high Galactic latitude ($|$b$|$ = 48$^\circ$) target $\epsilon$ Eridani.

\begin{table}
  \caption{Positions and flux densities of background sources. \label{table:bkg_fluxes}}
  \begin{tabular}{lccc}
  \hline
Source        & offset RA     & offset DEC  & Flux  \\
              &   [$\arcsec$] & [$\arcsec$] & [mJy/beam] 
  \\ \hline
Central peak  & ...           & ...     & 2.3 $\pm$ 0.3 \\
S1            &  85.9         &  20.0       & 4.6 $\pm$ 0.6 \\
S2            & -32.5         & -18.7 & 1.2 $\pm$ 0.3 \\
S3            & -34.5         &  -4.4 & 1.9 $\pm$ 0.3 \\
S4            & -52.0         &  11.4 & 3.0 $\pm$ 0.3 \\
S5            & -62.0         & -59.7 & 2.3 $\pm$ 0.4 \\
S6            & -73.1         &   5.6 & 4.2 $\pm$ 0.4 \\
S7            & -73.1         & -20.0 & 1.5 $\pm$ 0.3 \\
\hline
\end{tabular}
\begin{flushleft}Offsets are given with respect to the stellar position. At the time of the 
observations (2014.9) $\epsilon$ Eridani coordinates were $\alpha=$03:40:13.1 and 
$\delta=$-09:24:38.6 which include a correction for proper motion of 
$\mu_{\alpha}$=-975.2~$mas$/yr and $\mu_{\delta}$=19.5~$mas$/yr.\end{flushleft}
\end{table}

\section{Summary and conclusions}

We present the deepest (0.2~mJy rms) single dish observations at millimeter wavelengths of the 
prototypical debris disk target $\epsilon$ Eridani conducted with the AzTEC camera on the LMT. 
Our 7.5~arcmin$^{2}$ image reveals  the stellar emission, the cool disk, and
nearby (line-of-sight) environment with the following features: 

\begin{itemize}

\item The ring is detected for the first time at all position angles. The ring has a measured radius of 
20$\arcsec$ or 64~AU, and an upper limit of the width of 30~AU derived from model fitting, which 
implies a relative ring width 
($\Delta$R/R) of $\leq$ 0.5. These values are in agreement with previous observational and 
modeled properties. The ring shows some inhomogeneities that could be  explained by the presence of
background objects that coincide with the ring position. Bright structures in the ring previously observed at 
comparable sub-mm wavelengths with SCUBA, can also be ascribed to background objects, currently separated from the 
ring due to the star's high proper motion. Modeling of the ring indicates that its morphology is essentially smooth, 
and that a potential residual brightening in the SE might be an extension of the diffuse emission 
east from the system.

\item The central peak, which includes a stellar contribution and one or perhaps two warm dust belts
is also clearly detected. These LMT observations, along with the recent SMA data and archival {\it Herschel} 
fluxes at 70 and 160~$\mu$m, where the resolution is also good enough to separate the outer ring 
and the central peak, indicate that the interior warm dust contributes approximately 
60\% of this emission. The theoretical analysis of both the central peak and the outer dust ring shows evidence of 
significant emission in the gap. This may constitute the first evidence of  the Pointing-Robertson 
drag outside the solar system.

\item Numerous point sources are detected around the $\epsilon$ Eridani system. The sources most likely
correspond to a population of massive distant star-forming galaxies as has been suggested in previous works.
Further analyses are needed to verify the nature and the properties of their cool dust component.

\end{itemize}

Our 1.1~mm observations demonstrate the current capabilities of the operational LMT 
in the study of nearby circumstellar debris disks. 
We have traced the full extent of the nearest debris disk/Edgeworth-Kuiper belt analog 
to the solar system for the first time. At the same time we have identified a number of line-of-sight 
background sources, which could be members of the sub-mm bright, high redshift population of 
star-forming galaxies. Disentangling these extragalactic sources from excesses around 
disk-host stars is critical for the proper interpretation of planetary 
systems analogous to our own.  The LMT is expected to operate 
at its full aperture capacity of 50 meters in 2017. The resolution to be achieved in this final 
50-m diameter configuration of the LMT is 5$\arcsec$, and when combined with the increased 
sensitivity, stronger constraints on the ring properties will be observationally 
established.

\section*{Acknowledgments}
This work would have not been possible without the long-term financial support 
from the Mexican Science and Technology Funding Agency, CONACyT (Consejo Nacional de
Ciencia y Tecnolog{\'\i}a) during the construction and operational phase of 
the Large Millimeter Telescope Alfonso Serrano,
as well as support from the US National Science Foundation via the 
University Radio Observatory program, the Instituto
Nacional de Astrof\'isica, Optica y Electr\'onica (INAOE) and the
University of Massachusetts, Amherst (UMass). MC, EB, FCSM, MO and RLV work
was supported by CONACyT research grants SEP-2009-134985 and SEP-2011-169554. 
GMK is supported by the Royal Society as a Royal Society University Research Fellow. 
CE is partly supported by Spanish grant AYA2014-55840-P. JPM is supported by a 
UNSW Vice Chancellor's Postdoctoral Fellowship. SL acknowledges support from CONACyT 
through grant 238631. We are grateful to all of the LMT personnel and observers from 
Mexico and UMass who made possible this project. 

\bibliography{ms_epseri_final_mnras_revised}


\label{lastpage}

\end{document}